\documentclass[aps,prl,twocolumn,groupedaddress]{revtex4}
\usepackage{graphicx}
\begin{document}

%\preprint{}

\title{ Study of the $\eta\pi$ and $\eta'\pi$ spectra and 
 interpetation of possible  exotic $J^{PC}=1^{-+}$ mesons }



\author{ Adam P. Szczepaniak}
%\email[]{aszczepa@indiana.edu}
%\homepage[]{}
%\thanks{}
\affiliation{Physics Department and Nuclear Theory Center, 
Indiana University, Bloomington, 
Indiana 47405}
%\affiliation{Nuclear Theory Center, Indiana University, Bloomington, 
%Indiana 47405}

\author{ Alex R. Dzierba } 
%\email[]{dzierba@indiana.edu}
%\homepage[]{}
%\thanks{}
\affiliation{Physics Department, Indiana University, Bloomington, 
Indiana 47405}

\author{ Maciej Swat }
%\email[]{mswat@indiana.edu}
%\homepage[]{}
%\thanks{}
\affiliation{Physics Department and Nuclear Theory Center, 
Indiana University, Bloomington, 
Indiana 47405}
%\affiliation{Physics Department and Nuclear Thory Center, 
% Indiana University, Bloomington, 
%Indiana 47405}
%\affiliation{Nuclear Theory Center, Indiana University, Bloomington, 
%Indiana 47405}

\author{ Scott Teige }
%\email[]{teige@dustbunny.physics.indiana.edu}
%\homepage[]{}
%\thanks{}
\affiliation{Physics Department, Indiana University, Bloomington, 
Indiana 47405}


\date{\today}

\begin{abstract}
We discuss a coupled channel analysis of the $\eta\pi$ and $\eta'\pi$ systems 
produced in $\pi^-p$ interactions at 18~GeV/$c$.  
 We show that  known $Q\bar Q$ resonances, together with residual soft 
 meson-meson rescattering, saturate the spectra including the 
  exotic $J^{PC}=1^{-+}$ channel. A possibility of a narrow exotic resonance at a 
 mass near 1.6~GeV/$c^2$ cannot, however, be ruled out. 
\end{abstract}

%\keywords{}

\maketitle

 Gluons play a central role in the structure of matter as
  they mediate the strong 
 interactions leading to quark and gluon confinement.  
% This interaction is responsible for over $99\%$  of the baryonic mass
% in  the Universe with only the remaining $1\%$ due to the Higgs  mechanism.  
  The quantitative description of the dynamics of low energy gluons,
 however, is still unavailable and the direct manifestation of gluons in
the current spectroscopy of hadrons, for the most part, seems to be elusive.  Any theoretical
 approach is complicated by the strong character of the 
 interactions of soft gluons in QCD. Furthermore, the phenomenology of gluonic excitations 
 is challenging because gluons do not  carry electromagnetic or
 weak charge to which probes can directly couple. 
  The spectroscopy of hadrons with excited gluonic
 degrees of freedom remains as the only avenue to obtain information about 
 the nature of the soft gluonic modes. In this context 
 the spectroscopy of exotic mesons is of fundamental importance. 
 Exotic mesons have quantum numbers that cannot be attributed to valence
 quarks alone. Lattice gauge simulations indicate the gluonic
field confining quarks forms flux tubes and the quantum numbers of the
excited flux tubes couple with those of the quarks leading to
exotic quantum numbers. The  spin ($J$), parity ($P$) and charge conjugation ($C$)
 quantum numbers of the lightest exotic isovector should be
 $J^{PC}=1^{-+}$ and have mass around 1.9~GeV/$c^2$~\cite{l1,l2,l3}. It is
 possible, however, that corrections due to extrapolation of the
 lattice mass predictions  to the physical light quark masses 
 could introduce a 100 - 200~MeV/$c^2$ downward shift~\cite{astt}.  
  Based on the large $N_C$ expansion it has been
  shown that exotic mesons ought to have hadronic decay widths
 comparable to the other mesonic resonances~\cite{cohen} {\it i.e.} to
 be of the order of $\Gamma = $100 - 200~MeV/$c^2$.  Lattice
 simulations also indicate that at least one exotic meson multiplet
 should exist below the threshold energy for string breaking and quark 
 production~\cite{colin}. 

  Finding a $J^{PC}=1^{-+}$ wave 
  is not equivalent to establishing the existence of a QCD
  exotic.  A partial wave can have many poles as a function of complex
  energy in the unphysical domain of the complex energy plane
  (Riemann sheets). With an incomplete knowledge of the underlying
  strong interaction dynamics it is impossible to unambiguously 
 determine positions of distant poles. Similarly limited experimental
  data are unable to isolate and interpret such poles. 
  Only poles relatively close to the real axis can potentially be
  unambiguously identified, however, even in this case theoretical input
  is needed to discriminate between a preexisting bound state and
 a dynamical resonance.

The current experimental situation appears to be puzzling. Recently three exotic  
 $J^{PC} = 1^{-+}$  meson candidates have been reported.
  The $\pi_1(1400)$, which has an unusually light mass,
 $M= 1370 \pm 16^{+50}_{-30}  \mbox{ MeV}/c^2$, and a 
 large width, $\Gamma = 385 \pm 40^{+65}_{-105} \mbox{ MeV}/c^2$, was reported
  by  the E852 collaboration  in $\pi^- p \to \pi^-_1(1400) p \to
  \pi^-  \eta p$~\cite{eta} at 18~GeV/$c$. This state was  confirmed  in the
  $\eta\pi$ channels in    $ {\bar p} d$ annihilations to $\pi^- \pi^0 \eta p$  by the 
 Crystal Barrel  collaboration~\cite{cb}.   Two other $\pi_1(1600)$
  exotics  were also reported by the E852 collaboration,  one with $M =
  1597  \pm 10^{+45}_{-10}\mbox{ MeV}/c^2$ and  a large width of
   $\Gamma = 340 \pm 40 \pm 50\mbox{ MeV}/c^2$  observed in the
   $\eta'\pi^-$   channel~\cite{eprim}, and the other with 
 $M = 1593 \pm 8^{+29}_{-47} \mbox{ MeV}/c^2$ and a normal width of
   $\Gamma = 168 \pm{20}^{+150}_{-12} \mbox{ MeV}/c^2$ decaying into 
 $\rho^0\pi^-$~\cite{rho}. 
  The $\eta\pi$ and $\eta'\pi$ channels are well suited for
  $J^{PC}=1^{-+}$  exotic meson searches, since a neutral $P$-wave
  resonance has  exotic quantum numbers.  

 It should be noted that the above resonance parameters come from a Breit-Wigner
 (BW)  parameterization thus assuming a resonance
 interpretation of the data.  But a BW parameterization is only applicable if
 amplitudes  have singularities near the real axis, which in the case of a
  $300-400\mbox{ MeV}/c^2$ wide resonances is not reasonable. 
  Furthermore, in the original analysis of the E852 data,
   only selected $J_z$ states of the $\eta\pi$ and $\eta'\pi$ systems
  were considered in the mass dependent analysis that gave the 
  resonance parameters. If an exotic $J=1$
  resonance is produced, however,  all
   $2J+1$ spin components should be included in the analysis.  Even though these 
spin projections can have different
   production characteristics they  should all display similar resonance 
  behavior as a function of the invariant mass of the decay
  products. In a recent analysis of the $\eta\pi^0$ E852
 data~\cite{alex}  it was 
  shown that an exotic $P$-wave is indeed present, and very similar to
 the one
   found in the $\eta\pi^-$ mode, but when all information was used, 
 including all spin 
  components and production modes,  no self-consistent 
  set of BW parameters describing the observed $P$-wave could be found.

 Within QCD, the  quark structure of a typical narrow meson 
 resonance,  ${\it e.g}$ the $\rho(770)$ or the $a_2(1320)$,  
 is not much different 
 from a bound state. The quark wave functions are compact and the small
 width arises from coupling to a few open channels. In contrast,
 broad structures like the $\pi_1(1400)$ or the 
$\pi_1(1600)$, as observed in the $\eta\pi$ and $\eta'\pi$ channels,
 respectively, are likely to be of a different origin. In this paper
 we  
 present
 a new analysis of the E852 data and explain the dynamical origin of
 the $P$-wave enhancement seen in the $\eta\pi$ and $\eta'\pi$ channels.
 In particular, we  argue that these might be very 
  similar to the $\sigma$ meson which is used to parameterize the low
 energy $S$-wave,  isoscalar  $\pi\pi$ spectrum. It is generaly 
 accepted that the $\sigma$  meson does not originate from a 
 quark bound state but  is  instead a dynamical resonance arising from 
 residual, low energy $\pi\pi$ interactions~\cite{julich,oset}.

The low energy interactions of the pseudoscalar meson nonet 
 can be constrained  using standard low energy expansion methods, {\it
 e.g.} effective range theory. Chiral symmetry provides a
 natural scale for such an expansion, $f \sim 100\mbox{ MeV}/c$~\cite{gasser}. 
 In particular, for a two-body potential acting between two-meson
 channels, $i,j$ in $L$-th partial wave one has, 
 \begin{equation}
V^L_{ij}(p,q) = \langle p,i|V_L|q,j\rangle = (pq)^L \left[ c_{0,ij}
      + c_{2,ij}  {{p^2 +
      q^2} \over {f^2}} \cdots \right] 
\end{equation}
 with, $p$  and $q$ denoting the relative momenta between mesons. 
 It is expected that the coefficients,  $c_i$ are 
  $O(1)$, and that 
 these coefficients determine the low momentum expansion of
 the scattering amplitude. The real part of the scattering amplitude, 
 however, arises from iteration of the potential leading to
 important non-analytical  contributions from integration over 
 low momentum components of the  virtual states. 
 These non-analytical contributions have, in many cases, been found to be 
 sufficient  to extend the range of applicability of the low momentum 
 expansion  to regions well above the momentum scale set by $f$,   
even up to $1\mbox{ GeV}/c$~\cite{oset,tony,tony1}. 
  Furthermore, it is expected that higher terms in the expansion can be 
  approximated by pole terms corresponding to genuine  QCD bound
 states.  The approach can also be verified {\it a
 posteriori} by studying how sensitive are the parameters
 of these bound sates and intensity of the coherent backgrounds 
  introduced by $V_L$ to the details of the approximations. 
    In particular, we have studied the $N/D$ approach  
  with $V_L$ determining  the numerator and the poles determining the CDD
 terms~\cite{CDD}.  We also examined the Lippmann-Schwinger formalism with and
 without relativistic kinematics. Finally, in  all cases we studied
 various methods for removing the large momentum contributions from
 loop integrals, {\it e.g.} with or without explicit subtraction of
 power divergences. 
  Below we present results which we have found to be robust 
 with respect the various implementations.  
% More generally, the role of crossed channels
% and higher partial waves should be examined.

The distribution of events, $N(M_X, \Omega, t)$, 
 at fixed energy in the reaction 
 $\pi^- p \to X^- p$ with $X=\eta\pi^-$ or $\eta'\pi^-$, is
 proportional  to $|\sum_{LM} a_{LM}(t,M_X)  Y_{LM}(\Omega)|$.  
 The partial wave amplitudes, $a_{LM}$, determine the production
 strengths  of a   two-pseudoscalar meson system, with spin $J=L$, as a
 function  of their invariant mass ($M_X$) and the momentum transferred
squared, $t$, to the  target. The peripheral nature of the high-$s$,
 low-$t$ production implies that the $M_X$ and $t$ dependence is
 expected to factorize. Verification of the assumption can be found 
 in~\cite{alex}. The dependence on the solid angle, $\Omega$, 
 is defined with respect to the Gottfried-Jackson frame.  For a mass 
 dependent analysis we bin the data in $M_X$ and integrate
  over all $t$ (the limited statistics prevents us from considering more
 than two $t$-bins).  The number of events, $N(M_X,\Omega)$,
  in a given mass bin is decomposed into moments defined as $H_{LM}(M_X) = \int
  d\Omega N(M_X,\Omega)$ which are  corrected for detector
   acceptance.  The partial wave amplitudes are calculated from  
  two-meson interactions as described above and fitted to the moments. 
 In addition, it is necessary to specify the production amplitudes, $P^L_i$,
  whose mass dependence  can, in principle, be constrained by 
 duality. Instead, we use the requirement that they should
 have a  expansion similar to the potential,
  \begin{equation} 
 P^L_i(q) = \langle q,ip|V_L|\pi^-p \rangle = A^L_i q^L 
\left( 1 + d_{2,i}{{q^2}\over  {f^2}} \cdots \right)
 \end{equation}
 with the relative strengths satisfying, $A^L_i/A^{L'}_i = O(1)$. 
 In the mass range from threshold to $M_X<2\mbox{ GeV}/c^2$ we find that
 only the $L=S,P,D$ waves are needed to describe the data. 
 The partial waves are further classified by the magnitude of the
 spin projection, $M=0,1,\cdots L$ onto the beam axis and parity 
 under reflection in the production plane.  We find that only
 amplitudes with $M<1$ are relevant, leading to a  set of seven
 partial waves, $S,P_0,P_\pm,D_0,D_\pm$. Finally the
 positive and negative parity waves do not interfere and arise from
 $t$-channel natural and unnatural parity exchange, 
 respectively.

 For completeness, in addition to
 the $\eta\pi$ and $\eta'\pi$ channels, we also include the isosvector 
 $K\bar K$ channel. In the $S$-wave the $K\bar K$ attraction near
 threshold produces a cusp at $M_X \sim 1 \ \mbox{GeV}/c^2$ which corresponds
  to the $a_0(980)$ meson~\cite{oset}. A QCD resonance is expected
 at $M_X \sim 1.3 - 1.4 \mbox{ GeV}/c^2$ and could be associated with the 
 $a_0(1400)$ meson.  This is similar to the scalar-isoscalar
 channel where the low energy $\pi\pi$  attraction results in  the
 initial growth of the phase shift from threshold to $~700\mbox{ MeV}/c^2$
 which is commonly attributed to an effective $\sigma$ meson. 
  The opening of the attractive $K\bar K$ channel responsible for
 the $f_0(980)$ meson leads to  the rapid
  phase increase at $1\mbox{ GeV}/c^2$. Finally further growth of the phase in
  the $1.3 - 1.5\mbox{ GeV}/c^2$ can be attributed to presence of QCD
  resonances.  In the $S$-wave we thus include all three coupled
  two-meson channels and a single pole term with a mass expected 
 around  $1.3-1.4\mbox{ GeV}/c^2$~\cite{julich,oset}. 
  The $D$ wave is dominated by the narrow
 $a_2(1320)$ meson which we introduce as a pole term. The soft interactions
  in the $D$-wave correspond to $O(p^4/f^4)$ in the chiral expansion and
  thus cannot be unambiguously disentangled from pole contributions,
 also assumed to be $O(p^4/f^4)$. This introduces a theoretical
 uncertainty in the coherent $D$-wave background. Thus 
  the only constraint imposed on the $D$-wave is that coefficients are
 of $O(1)$.   The ability to correctly describe the 
 known spectrum in the $S$ and $D$-wave in an important check on both the
  theoretical method and data selection.

The $P$-wave is of the prime interest here. We have studied 
 cases with and without a pole term. The soft interactions in the
 $P$-wave channel are less constrained than, for example, the $S$-wave since
  they require mapping between the physical, $\eta,\eta'$ and the
  $SU(3)$  flavor, $\eta_0,\eta_8$  states~\cite{marco}. 
 In general it is predicted
 however that both $\eta\pi$ and $\eta'\pi$ interactions are attractive
  with the former weaker by $\sin^{2}(\theta) \sim 0.01$ 
 where $\theta$ is  the  $\eta_0 - \eta_8$ mixing angle.  
 With these general assumptions we have found that the 
 predictions for the spectrum of the $P$-wave is very insensitive to
 the details of the channel interactions.

 \begin{figure}
 \includegraphics[width=3.5in]{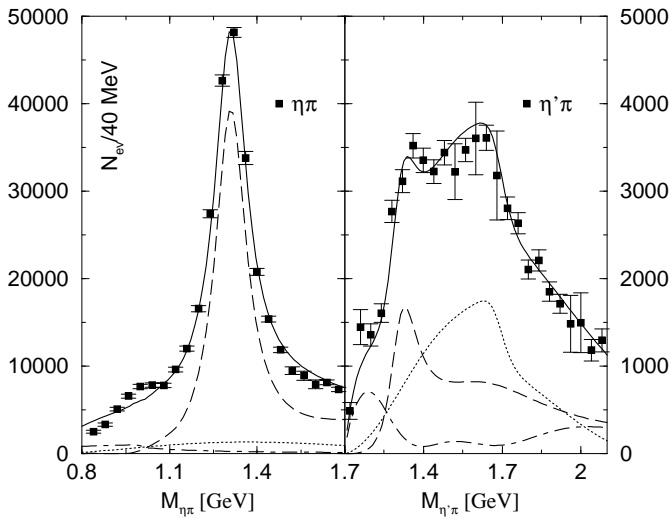}
 \caption{\label{fig1} Solid lines represent 
 $\eta\pi^-$ (left) and $\eta'\pi^-$ (right) mass
   spectra compared with the data. Also shown are 
 intensities of some of the partial waves, the $D_+$
   (dashed), the $S$ (dashed-dotted) and the $P_+$ wave (dotted). In
   the $\eta\pi$ channel the $S$ and $P$ (larger then $S$ at higher
   mass) are a small  fraction of the dominant $D$-wave.  For $\eta'\pi^-$ 
 the $D$  and $P$ waves are comparable.}
 \end{figure}

 \begin{figure}
 \includegraphics[width=3.5in]{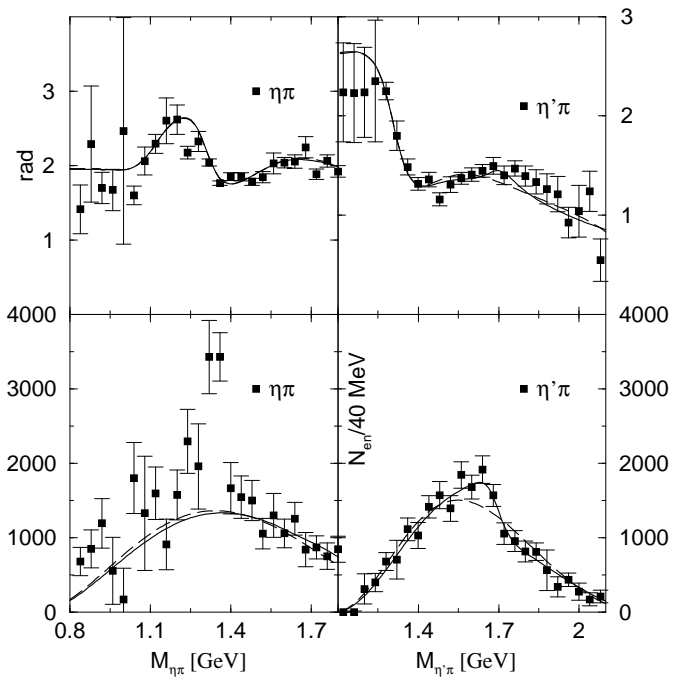}
  \caption{\label{fig2} Exotic $J^{PC}=1^{-+}$ wave coupled channel fit 
  to the $\eta\pi$  and $\eta'\pi$ data. The upper panels show  the
  phase difference (in radians) between $D_+$ and $P_+$ waves. The lower
  show the intensity of the $P_+$ wave. The dashed line 
  represents results of a fit without the pole term. }
 \end{figure}

 The results of the combined analysis of the $\eta\pi^-$ and $\eta'\pi$
 channels are shown in figures~1 and 2. The $\eta\pi$ channel is
 dominated by the $a_2(1320)$ resonance. The contribution from the
 coherent $D$-wave background is small leading to a pole mass within
 $10\%$ of the resonance mass listed by the PDG. The
 $S$-wave production is very weak 
  and does not constrain parameters of the $S$-wave
 pole, the $a_0(1400)$ not the dynamically generated, $a_0(980)$. 
 The discrepancy between data and theory in the $\eta\pi$ intensity 
 at $M_X \sim 1\mbox{ GeV}/c^2$ is within systematical errors of the
 $S$-wave parameters.

We did find, however, that the $H_{21}$ moment 
 which is sensitive to the interference between the unnatural
 exchange, $D_-$ wave and the $S$ wave, has nontrivial structure at 
 $M_X = 1.4\mbox{ GeV}/c^2$ which could be related to the the
 $a_0(1400)$ pole.  The $P$ wave intensity in the
 $\eta\pi$ channel is weak, $\le 5\%$ of the $D$ wave, and 
 leads to a broad structure consistent with the absence
 of nearby resonance in this channel. As discussed in ~\cite{alex}
 the peak-like structure at $M_X = 1.3~\mbox{GeV}/c^2$ is
 most likely due to the dominant $a_2(1320)$  leaking into the $P$-wave
 due to imperfect acceptance corrections. The phase of the natural
 exchange $P_+$ wave is found to  be very weakly mass dependent, 
 $|\delta \phi_P| < 10^0$ over the mass
 range considered, and as shown in figure~2(d) consistent with the data. The
 variations in the $\phi_D - \phi_P$ phase difference at $M_X> 1.6
 \mbox{ GeV}/c^2$ is due entirely to the weak coherent $D$-wave
 background. Similarly the sharp variation of this phase difference at
 low mass can be fully accounted for by the $D_+$ coherent
 background. Even though the channel interaction in the $D$ wave 
 are not well constrained it seems clear that both phase and magnitude of the
 $P$-intensity in the $\eta\pi$ channel can be well accounted for by
 $\eta\pi$ rescattering. In contrast, in the $\eta'\pi$ channel the 
 contribution 
 from the $P$ wave is significant and consistent with the theoretical
 expectations discussed above. As shown in figure~1b, the $P$ is as large as
 the $D$-wave while the $S$-wave is small. Since the $D$ and $P$
 waves are comparable there is no danger of one leaking into another and
 the partial wave analysis is quite unambiguous. It is seen from figure~2 that 
 the soft channel interactions describe  
 both the intensity and the phase motion of the $P$ wave very well.

We have 
 found, however, that there is a structure  near $M_X
  =1.6\mbox{ GeV}/c^2$ in several moments which could be accounted for 
 by a pole term in the $P$-wave. Since the bulk of the $P$-wave
  intensity 
 comes from soft re-scattering, the resonance associated with the
 pole term becomes narrow. If parameterized as a BW resonance it has 
 $\Gamma \sim 200~\mbox{MeV}/c^2$ and is thus consistent with the narrow
 resonance observed in the $\rho\pi$ channel.

In conclusion, we find that the broad $P$ wave in the $\eta\pi$ and
 $\eta'\pi$  spectra can be  accounted for by low energy
 re-scattering effects.  Comparing this  to the $\pi\pi$ channel, we
 interpret this enhancement as the equivalent  of the 
 $\sigma$ meson, {\it i.e.} arising from  correlated
  meson-meson state and not from a QCD bound state. 
 The applicability
 of low momentum expansion used here is well justified since, 
 after subtracting the threshold energies, the relevant 
 momentum range is comparable to the $\pi\pi$ case and the 
 two analyses are completely consistent.  In the non-exotic 
 $P$-wave $\pi\pi$ channel there are,  in addition to residual meson-meson
 interactions,   narrow resonances like the $\rho$ and $\phi$ mesons
 that are believed to be true QCD bound states.  
 We have found that the $\eta'\pi$ data does not rule out a narrow
 exotic $P$-wave resonance which might be the
 $\pi_1(1600)$  that is found to decay into $\rho\pi$ but from 
 no final conclusions can be drawn at this time. 

 This work was supported in part by the  U. S. Department of Energy.



\begin{thebibliography}{99}
%\end{document}
\bibitem{l1} P. Lacock {\it et al.}, UKQCD Collab., Phys. Rev. {\bf
    D}54, 6997 (1996). {\it ibid} Phys. Lett. {\bf B}401, 308 (1997).

\bibitem{l2} P. Lacock {\it et al.}, TXL Collab., Nucl. Phys. Proc. 
Suppl.{\bf 73},  261 (1999). 

\bibitem{l3} C.W. Bernard {\it et al.},  MILC Collab.,  Phys. Rev. {\bf
    D}56, 7039 (1997).

\bibitem{astt} A.W. Thomas, A.P. Szczepaniak, Phys. Lett. {\bf B}526,
  72 (2002). 

\bibitem{cohen} T.D. Cohen, Phys. Lett. {\bf B}427, 348 (1998). 

\bibitem{colin} K.J. Juge, J. Kuti, C.J. Morningstar,
  Phys. Rev. Lett. {\bf 82}, 4400 (1999). 



\bibitem{eta}
D.~R.~Thompson {\it et al.}  [E852 Collaboration],
Phys.\ Rev.\ Lett.\  {\bf 79}, 1630 (1997)

\bibitem{cb}
A.~Abele {\it et al.}  [Crystal Barrel Collaboration],
Phys.\ Lett.\ B {\bf 423}, 175 (1998).


\bibitem{eprim}
E.~I.~Ivanov {\it et al.}  [E852 Collaboration],
Phys.\ Rev.\ Lett.\  {\bf 86}, 3977 (2001)

\bibitem{rho}
G.~S.~Adams {\it et al.}  [E852 Collaboration],
Phys.\ Rev.\ Lett.\  {\bf 81}, 5760 (1998).


\bibitem{alex} A.R.~Dzierba {\it et al.} {\em 
 A Study of the $\eta \pi^{0}$ Spectrum and Search for a $J^{PC} =
 1^{-+}$  Exotic Meson}, {\it to appear in Phys. Rev. {\bf D}}, 
 arXiv:hep-ex/0304002. 

\bibitem{julich}
G.~Janssen, B.~C.~Pearce, K.~Holinde and J.~Speth, Phys.\ Rev.\ D {\bf
  52}, 2690 (1995)

\bibitem{oset} J.A. Oller, E. Oset, Phys. Rev. {\bf D}60, 074023
  (1999); J.A. Oller, E. Oset, J.R. Pelaez, Phys. Rev. {\bf
    D}59, 074001 (1999), Erratum-ibid. {\bf D}60, 099906 (1999). 

\bibitem{gasser} For a recent review see G. Colangelo, J. Gasser,
  H. Leutwyler, Nucl. Phys. B{\bf 603}, 125 (2001).



\bibitem{tony} A.~W.~Thomas, 
{\em Chiral extrapolation of hadronic observables}, 
 arXiv:hep-lat/0208023.

\bibitem{tony1} D.~B.~Leinweber, A.~W.~Thomas, K.~Tsushima and S.~V.~Wright,
Phys.\ Rev.\ D {\bf 64}, 094502 (2001)


\bibitem{CDD} L.~Castillejo, R.~Dalitz, F.~Dyson, Phys. Rev. {\bf
    101}, 543 (1956).

\bibitem{marco} S.D. Bass, E. Marco, Phys. Rev. {\bf D}65, 057503
  (2002). 


\end{thebibliography}
\end{document}